\documentclass{PoS}

\PoS{PoS(LAT2005)358}

\usepackage{epsfig}

\title{Chiral extrapolation of hyperon vector form factors\thanks{Preprint: ROMA-1414/05\,,\ \ RM3-TH/05-7\,.} }

\ShortTitle{Chiral extrapolation of hyperon vector form factors }

\author{D.~Guadagnoli$^a$,\ V.~Lubicz$^{b,c}$,\ G.~Martinelli$^a$,\
		M.~Papinutto$^d$,\ S.~Simula$^{c}$,\ \speaker{G.~Villadoro}$^a$\\
        {}$^a$~Dipartimento di Fisica, Universit\`a di Roma ``La Sapienza'',\\ 
  			\quad and INFN, Sezione di Roma, P.le A. Moro 2, I-00185 Rome, Italy \\
		{}$^b$~Dip. di Fisica, Universit\`a di Roma Tre, Via della Vasca Navale 84, I-00146 Rome, Italy \\
		{}$^c$~INFN, Sezione di Roma III, Via della Vasca Navale 84, I-00146 Rome, Italy \\
		{}$^d$~John von Neumann-Institut f\"ur Computing NIC, Platanenallee 6, D-15738 Zeuthen, Germany \\
        E-mail: \email{giovanni.villadoro@roma1.infn.it}
}

\abstract{We present a new study of SU(3)--breaking corrections in hyperon vector form factors relevant
for the extraction of $V_{us}$. A lattice quenched simulation has been performed, showing that it is
possible to reach the required precision to extract SU(3)--breaking corrections in the regime
of simulated masses. In order to perform the chiral extrapolation we calculated the 
chiral corrections to the vector form factor in HBChPT. Besides the one-loop $O(p^2)$ contribution, 
we included also the subleading $O(p^3)$ and $O(1/M_B)$ corrections that, due to the Ademollo-Gatto theorem, 
are free from the contamination of unknown low energy constants. 
The results complete and correct previous calculations, and show that subleading corrections 
cannot be neglected. We also studied decuplet contributions within HBChPT and show that, in this
case, the chiral expansion breaks down, rising doubts on the consistency of the theory.
}

\FullConference{XXIIIrd International Symposium on Lattice Field Theory\\
                 25-30 July 2005\\
                 Trinity College, Dublin, Ireland}

\begin{document}
\section{Introduction}
Recently, it has been shown that SU(3)-breaking corrections for 
vector form factors (v.f.f.) can be extracted from lattice simulations with a great precision \cite{kl3}.
The method of ref.~\cite{kl3} allowed to reach the percent level accuracy in the 
extraction of $V_{us}$ from $K_{\ell3}$ decays, stimulating new unquenched studies
to reduce systematic errors \cite{ukl3}. An independent way to extract $V_{us}$ is provided
by hyperon semileptonic decays. Ref.~\cite{csw} showed that, analogously to
the mesonic case, it is possible to extract the product $|V_{us} \cdot f_1(0)|^2$ at the 
percent level from experiments, with the v.f.f. $f_1(0)$ defined by ($q= p_1-p_2$):
\begin{equation}
\langle B_2 | V^\mu| B_1\rangle=\overline B_2(p_2) \left [ \gamma^\mu f_1(q^2) 
	- i\frac{\sigma^{\mu\nu}q_\nu}{M_1+M_2} f_2(q^2) +\frac{q^\mu}{M_1+M_2} f_3(q^2)\right ] B_1(p_1) \,.
\end{equation}
The Ademollo-Gatto (AG) theorem \cite{ag} protects $f_1(0)$ from linear SU(3)--breaking corrections that
are thus suppressed. Although experiments seem to be consistent with negligible SU(3)--corrections
\cite{csw}  they are not accurate enough to exclude sizeable effects (i.e. larger than percent)  
in the extraction of $V_{us}$. Model dependent estimates based on quark models, $1/N_c$
and chiral expansions give different results (see e.g. \cite{pich}) so that the 
lattice seems the right tool to address this problem. 
We completed the preliminary study of ref.~\cite{gmps} showing that it is indeed possible to
extract SU(3)--breaking corrections from the lattice with the method of ref.~\cite{kl3}.
One of the main sources of uncertainties in lattice simulations is the chiral extrapolation,
expecially for v.f.f. where the AG theorem makes these quantities
dominated by mesonic loops. We performed a systematic calculation of these corrections within
Heavy Baryon Chiral Perturbation Theory (HBChPT) \cite{jm-1} including 1--loop $O(p^2)$ corrections
as well as subleading $O(p^3)$ and $O(1/M_B)$ contributions. As for the mesonic sector, AG suppresses
contributions from counterterms at $O(p^4)$ and makes the corrections finite and free
from unknown parameters. They are real predictions of the theory and can be used therefore also to test the
convergence of the perturbative expansion. 
This analysis completes (and corrects) the calculations of ref.\cite{krause} and 
\cite{al}. We show that the convergence of the series is rather poor. We also
tested the inclusion of decuplet contributions which are expected to give important effects. We
find, however, that they seem to spoil completely the chiral expansion, raising strong doubts on the 
consistency of the theory itself. 
\section{Quenched Lattice results}
\begin{figure}[t]
\epsfig{file=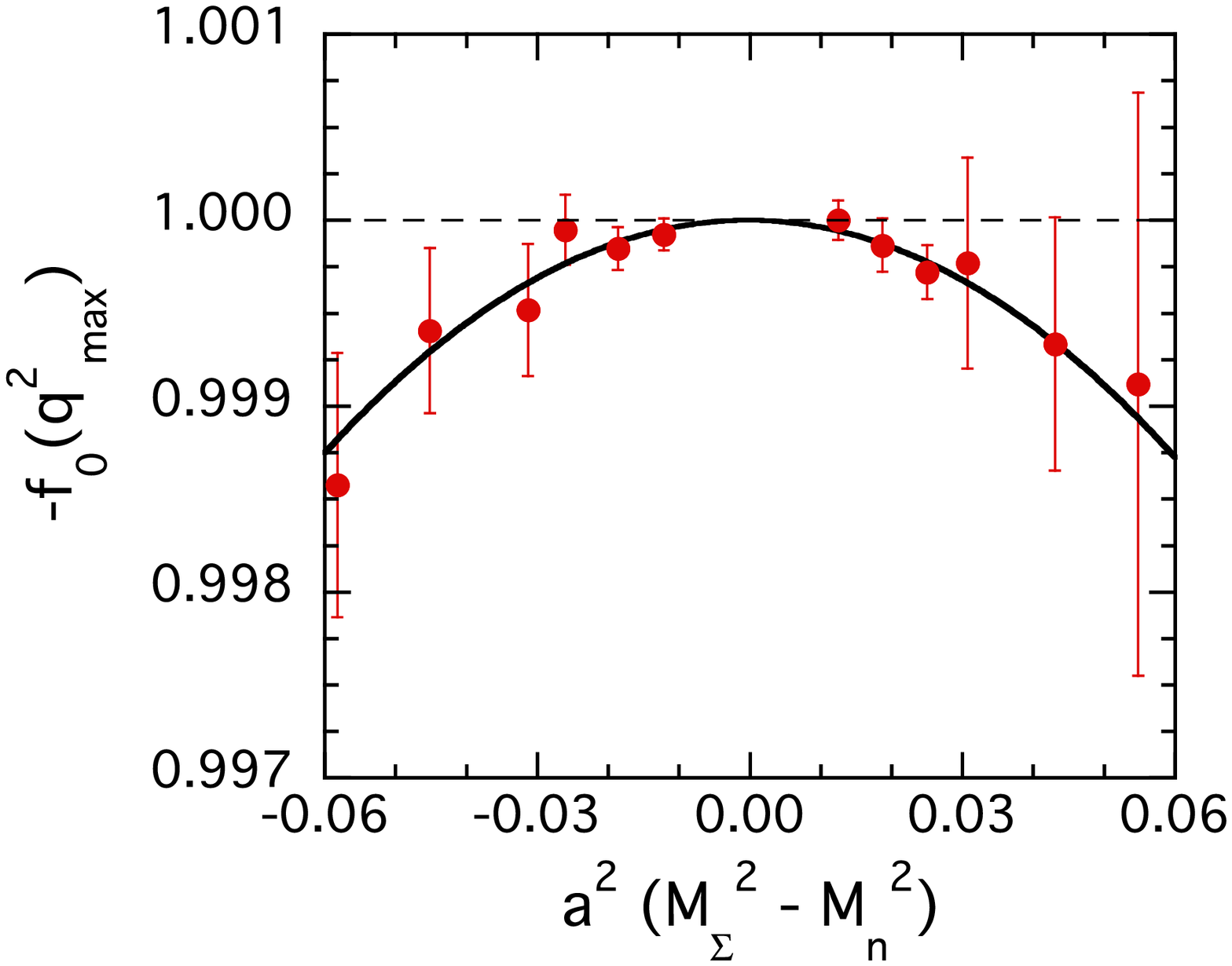,width=.39\textwidth} 
\epsfig{file=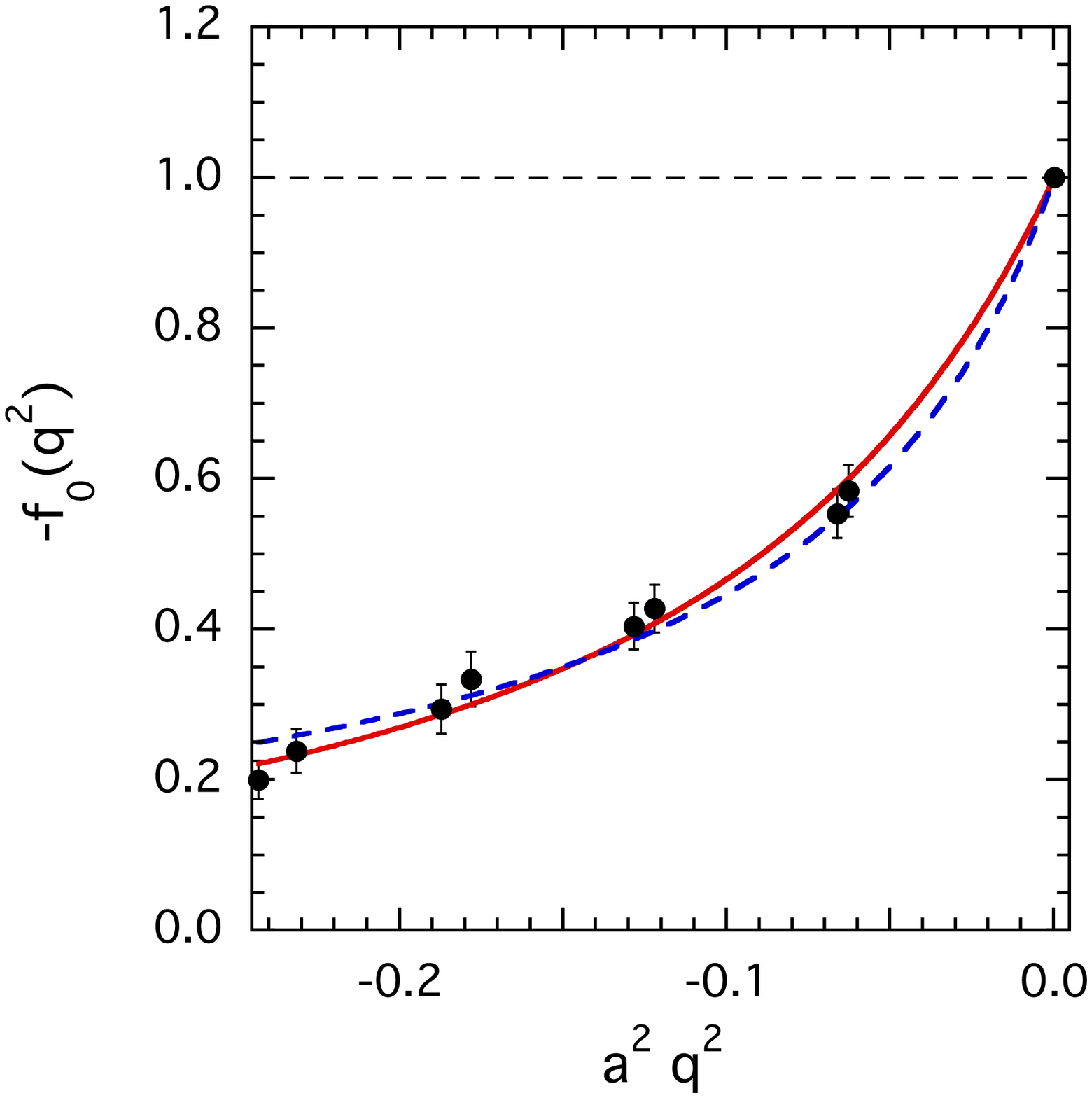,width=.30\textwidth}
\epsfig{file=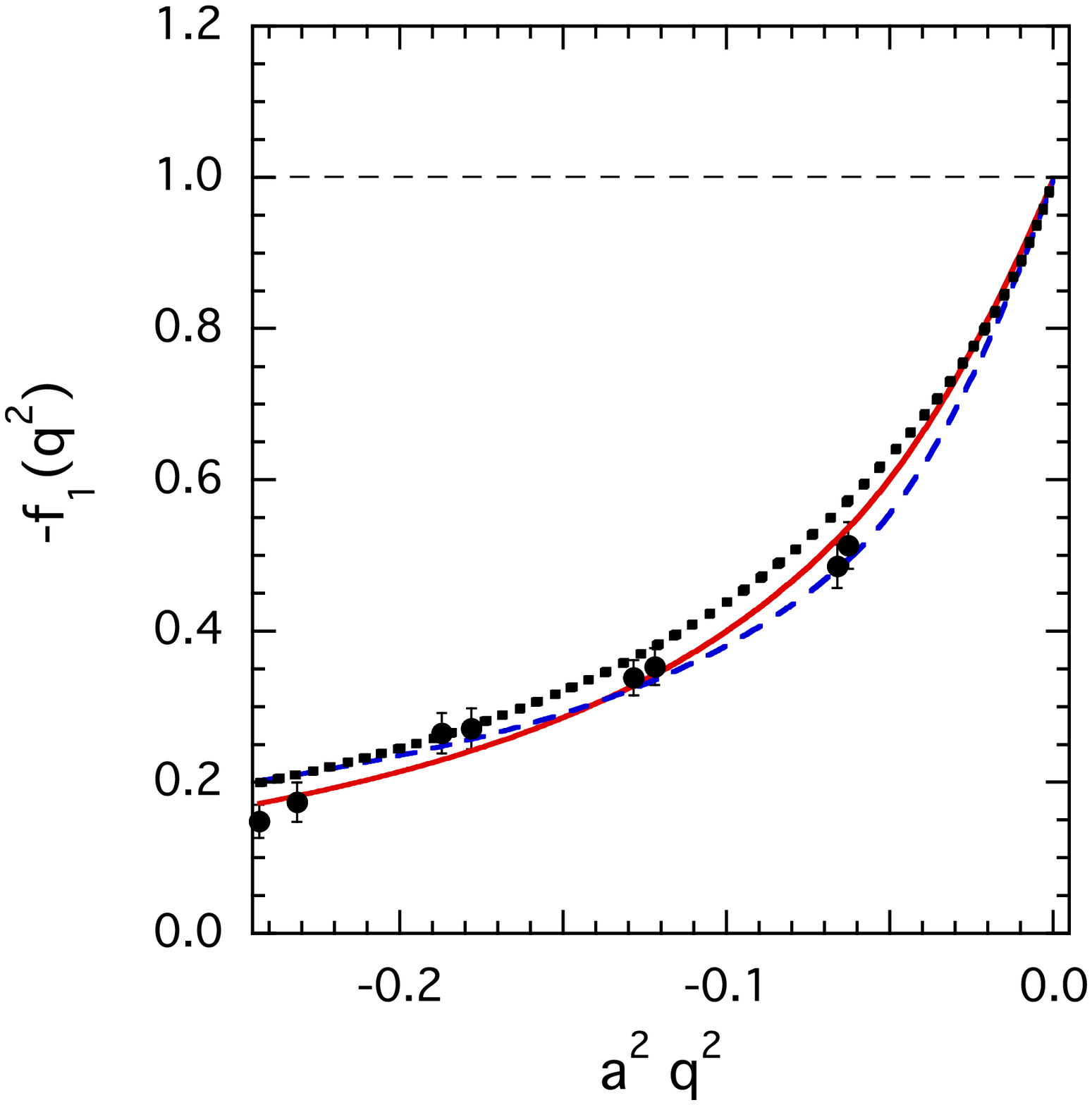,width=.30\textwidth}
\caption{The first plot shows the results for $f_0(q^2_{max})$ extracted from the double ratio (2.2). 
The other two plots show the fit in $q^2$ for $f_0$ and $f_1$ respectively. Curves correspond to
monopole fit (dashed blue), dipole fit (solid red) and dipole fit with fixed slope $\lambda=1/M_{K^*}^2$ (dotted black).}
\label{fig:q-dep}
\end{figure}
\begin{figure}[t]
\epsfig{file=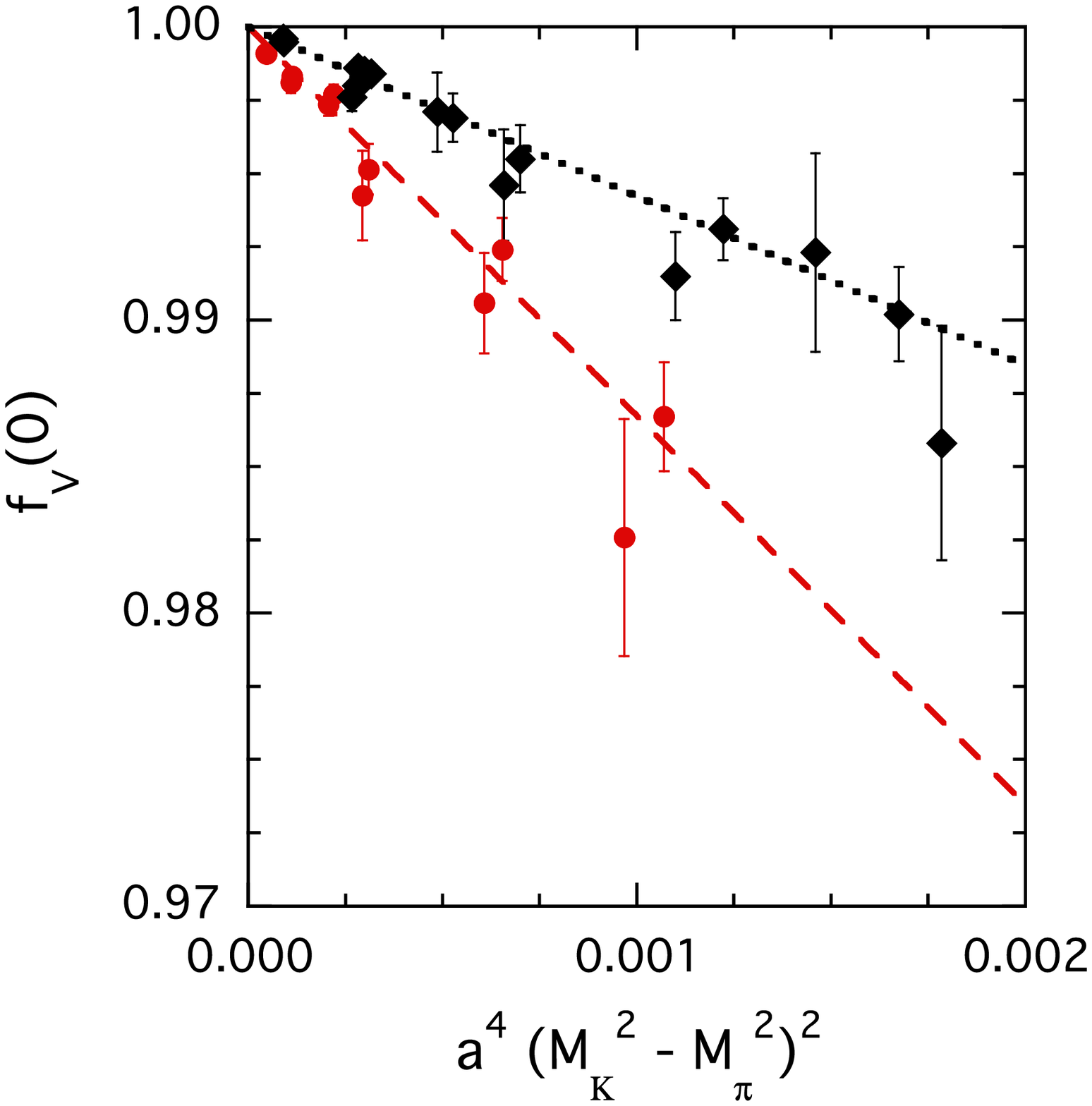,width=.45\textwidth,height=5.5cm}
\epsfig{file=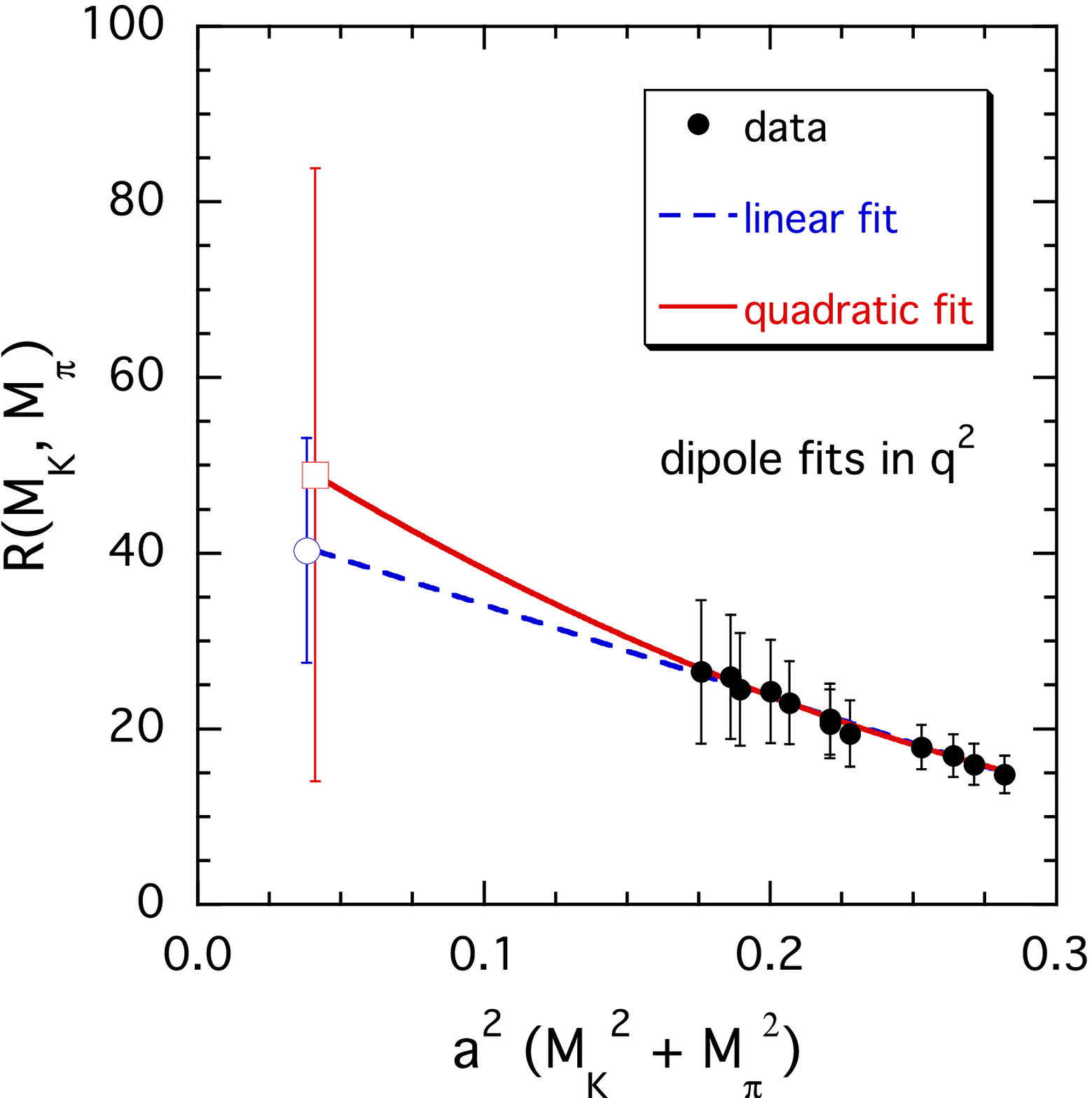,width=.45\textwidth,height=5.5cm}
\caption{Left: fit of "$-f_1(0)$" for $\Sigma\to n$ (dashed red)
compared to that of "$f_+(0)$" for $K\to\pi$ of ref.~\cite{kl3} (dotted black). 
Right: chiral extrapolation for the AG ratio $R$.}
\label{fig:masses}
\end{figure}

The lattice analysis is based on 240 quenched configurations with $\beta=6.20$ ($a^{-1}\simeq 2.6$~GeV) 
on a $24^3\times 56$ lattice and with quark masses corresponding to baryon masses in the range $M_{B}\sim (1.5\div 1.8)$~GeV. 
We considered $\Sigma^-\to n$  transitions, closely following the three step procedure described in \cite{kl3}. 
The first step consists in extracting the scalar form factor
\begin{equation}
f_0(q^2)\equiv f_1(q^2)+\frac{q^2}{M_\Sigma^2-M_n^2} f_3(q^2)\,,
\end{equation}
at $q^2_{max}=(M_\Sigma-M_n)^2$ via the Fermilab double ratio method \cite{fermilab}:
\begin{equation} \label{eq:doubleratio}
|f_0(q^2_{max})|^2 = \left [\frac{\langle n | \overline u \gamma_4 s | \Sigma^- \rangle 
	\langle \Sigma^- | \overline s \gamma_4 u | n \rangle}{\langle n | \overline u \gamma_4 u | n \rangle 
	\langle \Sigma^- | \overline s \gamma_4 s | \Sigma^- \rangle}\right ]_{\vec p_1 =\vec p_2=0}\,.
\end{equation}
This allows to obtain $f_0(q_{max}^2)$ with a very high precision (Fig.~\ref{fig:q-dep}). 
The second step is the study of the momentum dependence through other suitable double ratios
(see \cite{kl3}). Results for both $f_0(q^2)$ and $f_1(q^2)$ with dipole and monopole fits 
are reported in Fig.~\ref{fig:q-dep}. We checked that the slope from the dipole fit of $f_1$ is in agreement
with the experimental value $\lambda_1\simeq 1/M_{K^*}^2$. From these fit we extracted the values
of $f_0(0)=f_1(0)$ for different quark masses. Fig.~\ref{fig:masses} shows how they nicely agree with
the AG prediction. We can thus construct the AG ratio:
\begin{equation}
R(m_K,m_\pi)\equiv \frac{1+f_1(0)}{(a^2 m_K^2-a^2 m_\pi^2)^2}\,,
\end{equation}
which is found to depend mainly on $a^2 (M_K^2+M_\pi^2)$. Finally the plot of the chiral extrapolation 
of $R$ to the physical point is reported in Fig.~\ref{fig:masses}. Without a better knowledge of the 
chiral corrections we performed a linear and a quadratic fit that give, upon averaging, 
the extrapolated value:
\begin{equation} \label{eq:latres}
f_1(0)=-1+(5.3\pm3.8)\%.
\end{equation} 
The large uncertainty is mainly due to the long chiral extrapolation. Smaller quark masses should sensibly reduce it. 
However this result cannot be considered complete since we know that this quantity is dominated by meson (quark) loops
that, expecially in the quenched case, are not correctly taken into account by the lattice simulation. 
Either (almost) physical quark masses are used or chiral corrections have to be included.
\section{Chiral corrections in HBChPT}
In order to study the chiral behaviour of v.f.f. we use the HBChPT
formulated in \cite{jm-1} where baryons are treated as heavy degrees of freedom
and a $1/M_B$ expansion around the non-relativistic limit is performed. The chiral corrections
to the v.f.f. $f_{1}(0)$ can be schematically expressed as:
\begin{equation} \label{eq:expansion}
f_{1}(0)=f_1(0)^{SU(3)} \left \{ 1 + O\left ( \frac{m_K^2}{(4\pi f_\pi)^2}\right ) +\left [
O\left ( \frac{m_K^2}{(4\pi f_\pi)^2}\frac{\pi \delta M_B}{m_K} \right ) + 
O\left ( \frac{m_K^2}{(4\pi f_\pi)^2}\frac{\pi \,m_K}{M_B}\right )  \right ] + O(p^4)\right \}.
\end{equation}
$f_1(0)^{SU(3)}$ is the value of the v.f.f. in the SU(3) limit that is fixed by 
the vector current conservation (see tab.~\ref{tab:results}). The first term is the one-loop $O(p^2)$
correction and the terms in square brackets are the $O(p^3)$ and $O(1/M_B)$ corrections respectively.
These corrections should be suppressed with respect to $O(p^2)$ but the presence of a factor 
$\pi$, due to a double pole structure, makes these corrections numerically important. Two papers
already investigated these corrections but they did not perform a full calculation. 
The first one \cite{krause} used BChPT to $O(p^2)$ and did neither include $O(p^3)$ corrections 
nor take into account that, in general, the naive relativistic approach breaks the power counting.
The second \cite{al} used HBChPT including some of the $O(p^3)$
corrections but did not consider $O(1/M_B)$ corrections. Moreover there is a mistake
in the $O(p^2)$ corrections (a sign does not agree with \cite{krause}) and probably also in the $O(p^3)$ one.
It seems thus mandatory to redo the whole calculation. For shortage of space we will present all the explicit
formul\ae\ in a forthcoming paper. 

The first non trivial correction comes from one-loop graphs (see ref.~\cite{al}). Because
of the AG theorem, as for the $K\to\pi$ case, it can be expressed only in term of the AG preserving
function
\begin{equation}
H_{1,2}=m_1^2+m_2^2-2\frac{m_1^2m_2^2}{m_2^2-m_1^2}\log{\frac{m_2^2}{m_1^2}}\,,
\end{equation}
that is scale independent. There are two type of contributions at this order: tadpole and sunset. 
The former is a universal contribution which is the same for all the channels and equals that of
$K^0\to\pi^-$ ($-2.3$\%)\footnote{The sign of this correction agrees with \cite{krause} but disagrees with \cite{al}.}.
The latter, on the other hand, depends on the channel and on the two tree-level
axial couplings $D$ and $F$ that are well known ($D=0.80$, $F=0.46$, see ref.~\cite{csw}).
These corrections are of order $\pm(2\div 7)$\% (see tab.~\ref{tab:results})
and agree with both \cite{krause} and \cite{al}. 

$O(p^3)$ corrections are obtained by inserting $O(p^2)$ operators into one-loop diagrams. 
In the $O(p^2)$ HBChPT Lagrangian, there are many operators with unknown low energy constants (LECs).
We checked, however, that in $f_1(0)$ only those shifting the baryon masses can contribute. This fact allows to give
an estimate of the full $O(p^3)$ corrections that is free from the uncertainty due to the ignorance
of the LECs. The insertion of the baryon mass-shifts produces double poles in one-loop diagrams
and thus the $\pi$ factor in eq.~(\ref{eq:expansion}). We checked that our corrections agree with the AG theorem 
that represents a strong cross-check for the result. Notice that, since at this order baryons
are not degenerate anymore, the calculation does not correspond to the trivial insertion of baryon mass-shifts on
$O(p^2)$ diagrams, where both the external particles can be taken at rest. This might explain
the disagreement of \cite{al} with our results. The $O(p^3)$ corrections depend on $F$ and $D$ and 
give important positive contributions of order $2\div 6$\% (see tab.~\ref{tab:results}).

Finally, since $M_B\sim 1$~GeV, $O(1/M_B)$ corrections are $O(p)$ and their inclusion in one-loop diagrams gives
contributions of $O(p^3)$. However the coefficients of the $O(1/M_B)$ operators are fixed by Lorentz symmetry,
so that no new unknown LEC is introduced. We find that these corrections are important ($3\div 8$\%),
depend on $F$ and $D$ and tend to cancel the sunset part of the $O(p^2)$ corrections. They agree with the AG theorem and 
with the expansion to $O(1/M_B)$ of the result of \cite{krause}. 
We also notice a strong dependence on $M_B$ which represents the signal 
that higher order corrections could be also important.
\begin{table}
\begin{center}
\begin{tabular}{|| c || c |  c | c | c || c ||}
\hline 
$f_1(0)/f_1(0)^{SU(3)}$ & $f_1(0)^{SU(3)}$ & $O(p^2)$ & $O(p^3)$ & $1/M_B$ & All \\ \hline \hline
$\Sigma^-\to n$ & $-1$ & $+0.7$\% & $+6.5$\% & $-3.3$\% & $+3.9$\% \\ \hline
$\Lambda\to p$ & $-\sqrt{3/2}$ & $-9.4$\% & $+4.2$\% & $+8.2$\% & $+3.0$\% \\ \hline
$\Xi^-\to \Lambda$ & $\sqrt{3/2}$ & $-6.2$\% & $+6.1$\% & $+4.5$\% & $+4.4$\% \\ \hline
$\Xi^-\to \Sigma^0$ & $1/\sqrt2$ & $-9.1$\% & $+2.3$\% & $+7.9$\% & $+1.2$\% \\ 
\hline 
\end{tabular}
\end{center}
\caption{Chiral corrections at the physical point (physical masses, decay constants and 
axial couplings), $D=0.80$, $F=0.46$ and $M_B=1.1$~GeV}
\label{tab:results}
\end{table}
\section{Decuplet contribution and discussion}
The sum of all contributions for the different channels are reported in tab.~\ref{tab:results}. They are positive
and smaller than previously claimed in ref.~\cite{al}\footnote{Notice that ref.~\cite{al} used sensibly smaller
values for $D$ and $F$.}. Results in tab.~\ref{tab:results} are clearly not the final answer. Higher order 
corrections are expected to give large contributions as in the $K\to\pi$ case (\cite{lr,kl3}). Another
source of uncertainty is represented by the decuplet contributions in the effective field theory calculation.
In the decoupling limit, where the decuplet-octet mass-shift $\Delta$ is taken much larger than the interaction scale
$\Lambda_{QCD}$, decuplet contributions can be reabsorbed into the LECs and do not give any observable correction
at this order in the chiral expansion (notice that, by using physical values for masses and couplings, much of their
contribution is already taken into account). However $\Delta\simeq 230~{\rm MeV}\sim \Lambda_{QCD}$ and the 
decuplet might give non negligible non-analytic contributions to the chiral expansion. 
The HBChPT with explicit decuplet d.o.f. was firstly proposed in \cite{jm-2}, and formalised as an expansion in \cite{hhk}.
We used this approach to calculate the decuplet effects on $\Sigma^-\to n$ transitions. As for the octet contributions, 
the AG theorem protects the corresponding decuplet corrections from unknown LECs and the only new parameter, besides $\Delta$, is
the known decuplet--octet--meson coupling ${\cal C}\simeq 1.5$. 
At $O(p^2)$ the dynamical decuplet gives an important contribution ($-2.6$\%). At $O(p^3)$ there are two contributions.
The first is due to the insertion of decuplet mass-shifts and it is of order $-1$\%.
The second is due to baryon mass-shifts insertions and gives a contribution of order $-32$\%!
The large contribution with respect to the baryon's one could be explained by the stronger coupling
of decuplet to mesons, ${\cal C}^2/D^2\sim 4$. However this cannot explain why $O(p^3)$ decuplet
corrections are one order of magnitude larger than those of $O(p^2)$. This effect actually breaks the chiral expansion
raising serious doubts on the consistency of the HBChPT with the decuplet. The reason why this effect was not
noticed before is because other quantities, at this order, contain a large number of LECs that can
be adjusted to fit the data. In this case there are no LECs and a true test of the convergence of the
chiral expansion becomes possible. 
For this reason a model independent estimate of chiral corrections for baryons cannot be given at the moment.
The best we can do is to restrict ourselves to HBChPT without dynamical decuplet, making the ansatz that
decuplet contributions, though important, can be reabsorbed into local terms. Under this assumption
there seems to be a sort of cancellation between loop-corrections to $\Sigma\to n$ (tab.~\ref{tab:results}) 
and local contributions from the quenched simulation (\ref{eq:latres}).
However, without a better control on the theory, unquenched simulations with light quark masses,
which do not rely on the chiral expansion, are needed for a reliable estimate of hyperon form factors.

%
\end{document}